\documentclass[10pt,twocolumn,letterpaper]{article}
\pdfoutput=1
\usepackage{iccv}
\usepackage{times}
\usepackage{epsfig}
\usepackage{graphicx}
\usepackage{amsmath}
\usepackage{amssymb}
\usepackage{multirow}
\usepackage{booktabs}
\DeclareMathOperator*{\argmax}{arg\,max}

\usepackage{algorithm}
\usepackage{algorithmic}
\usepackage{xr}
\usepackage[accsupp]{axessibility}

\usepackage{color}

\usepackage[pagebackref=true,breaklinks=true,letterpaper=true,colorlinks,bookmarks=false]{hyperref}

\iccvfinalcopy 

\def\iccvPaperID{11013} 

\ificcvfinal\fi

\begin{document}

\title{Sample Efficient Detection and Classification of Adversarial Attacks \\ via Self-Supervised Embeddings}

\author{Mazda Moayeri\\
Department of Computer Science \\
University of Maryland\\
{\tt\small mmoayeri@umd.edu}
\and
Soheil Feizi\\
Department of Computer Science \\
University of Maryland\\
{\tt\small sfeizi@cs.umd.edu}
}

\maketitle
\ificcvfinal\thispagestyle{empty}\fi

\begin{abstract}
    Adversarial robustness of deep models is pivotal in ensuring safe deployment in real world settings, but most modern defenses have narrow scope and expensive costs. In this paper, we propose a self-supervised method to detect adversarial attacks and classify them to their respective threat models, based on a {\it linear} model operating on the embeddings from a pre-trained self-supervised encoder. We use a SimCLR encoder in our experiments, since we show the SimCLR embedding distance is a good proxy for human perceptibility, enabling it to encapsulate many threat models at once. We call our method {\bf SimCat} since it uses \underline {Sim}CLR encoder to \underline{cat}ch and \underline {cat}egorize  various types of adversarial attacks, including $\ell_p$ and non-$\ell_p$ evasion attacks, as well as data poisonings. The simple nature of a linear classifier makes our method efficient in both time and sample complexity. For example, on SVHN, using only five pairs of clean and adversarial examples computed with a PGD-$\ell_{\infty}$ attack, SimCat's detection accuracy is over 85\%. Moreover, on ImageNet, using only 25 examples from each threat model, SimCat can classify eight different attack types such as PGD-$\ell_2$, PGD-$\ell_\infty$, CW-$\ell_2$, PPGD, LPA, StAdv, ReColor, and JPEG-$\ell_\infty$, with over 40\% accuracy. On STL10 data, we apply SimCat as a defense against poisoning attacks, such as BP, CP, FC, CLBD, HTBD, halving the success rate while using only twenty total poisons for training. We find that the detectors generalize well to unseen threat models. Lastly, we investigate the performance of our detection method under adaptive attacks and further boost its robustness against such attacks via adversarial training.

\end{abstract}

\section{Introduction}

Deep learning has been applied to many applications with great success, though a major roadblock to safe deployment of deep models in real world settings is their susceptibility to adversarial attacks. Targeting a model at the time of inference is known as an evasion attack \cite{szegedy2013}, where imperceptible perturbations are made to an input to craft an \textit{adversarial example} that is misclassified by the model. Models can also be attacked during training via poisoning \cite{goldblum2020dataset}, where a small number of adversarial examples are inserted to a training set, so that after training, targeted test samples are misclassified. 

To mitigate these vulnerabilities, many defenses have been introduced \cite{madry2018towards,tramer2018ensemble,cohen2019certified,laidlaw2021perceptual}. However, a number of practical challenges remain. First, most defenses only harden a model against a specific narrow threat model \cite{kang2019robustness}. Thus, an attacker can easily evade the defense by using a different attack. Second, novel attacks are introduced frequently, in a brisk game of cat and mouse, in which the attacker generally has the upper hand \cite{athalye2018obfuscated}. When we combine the rapid development of attacks with the high computational cost of retraining a model to be robust against even a single threat, maintaining the robustness of a model to all threats using typical defenses becomes intractable. 

Certain detection based defenses \cite{Survey} do not require fundamentally changing how the model is trained, which allows for easy application of the detector without disrupting the existing machine learning pipeline. However, many detection based systems require large amounts of data. This is problematic in practice, as an adversary may employ an attack that the defender has never seen before, making it difficult for the defender to have trained the detector to work against it. Even if the defender obtains some examples of the novel attack, the training size may not be sufficient to expand the detector's ability to include the novel attack. 

The advantage of a detector that is both broad in scope and inexpensive in training requirements is clear. In this paper, we propose a highly sample efficient detector that also generalizes well to unforeseen attacks. Further, we extend our model to classify adversarial attacks to their respective threat models, similar to \cite{maini2021perturbation}. The classification allows for additional defenses to be employed, specific to the attack encountered. Our model is based on pretrained self-supervised encoders, which are capable of efficiently extracting the semantic content of images, as evident by recent successes in the self-supervised domain that have closed the gap with supervised models. 

Specifically, we use a SimCLR encoder pretrained on ImageNet, because we observe that distance in the embedding space of this encoder correlates strongly with human perception (Figure \ref{fig:percep_corr}). Furthermore, the distance between the SimCLR representations of a clean image and its adversarially perturbed counterpart is similar across various threat models, making the SimCLR distance a strong candidate to proxy the true perceptual threat model, which encapsulates all imperceptible adversarial attacks \cite{laidlaw2021perceptual}. While LPIPS \cite{zhang2018perceptual} has also been shown to be a strong proxy for true perceptual distance, the dimensionality of its feature vectors is massive in comparison to that of SimCLR representations. Our method is called {\bf SimCat}, as it uses a SimCLR encoder to linearly \underline{cat}ch adversarial attacks and \underline{cat}egorize them to their respective threat models.


By utilizing the highly informative low dimensional embedding space of self-supervised encoders, we find that even a linear model trained on these representations can effectively detect and classify adversarial attacks of a wide variety of types. Our experiments over many threat models and multiple datasets show that, for both evasion and poisoning attacks, SimCat greatly outperforms baseline models. While the method is extremely simple, we argue that its simplicity is essential for the efficiency of the method both in terms of time and sample complexities. By freezing the encoder, SimCat's optimization is \textit{convex} and thus its global optimum can be found effectively. Moreover, SimCat has low model complexity due to the small dimensionality of SimCLR representations, allowing for highly efficient training that also generalizes well. These properties lead to an impressive empirical performance of SimCat in detection and classification of various types of adversarial examples using as few as 5 training samples per class.

For example, on ImageNet, using only 5 samples per threat model, SimCat's detection accuracy is 68.5\%, improving the baseline method by more than 6.4\%. Using the same setup, SimCat's classification accuracy of 8 adversarial attacks including PGD-$\ell_2$, PGD-$\ell_\infty$, CW-$\ell_2$ \cite{carlini2017adversarial}, PPGD \cite{laidlaw2021perceptual}, LPA \cite{laidlaw2021perceptual}, StAdv \cite{xiao2018spatially}, ReColor \cite{laidlaw2019functional}, and JPEG-$\ell_\infty$\cite{kang2019robustness} is 27.1\%, improving the baseline performance by 7.7\%. Using 25 samples per class, SimCat's gains over baseline method grows to 7.4\% and 11.8\% for detection and classification problems, respectively.

Interestingly, SimCat can be used to detect and classify various types of poisoning attacks as well, which we then apply as an efficient poison defense. We consider five types of poisonings including bullseye polytope (BP) and convex polytope (CP) \cite{aghakhani2021bullseye, zhu2019transferable}. Using two SimCat detectors trained with 10 BP and 10 CP poisons respectively, we construct an ensemble detector to remove any sample that is flagged as poison by either of the individual detectors. The ensemble reduces poison success rate over five types of poisoning attacks from 21.8\% to 9.7\%. Notably, the SimCat poisoning defense only reduces clean accuracy by 1\%. 

Lastly, we develop an adaptive attack that creates adversarial examples to evade SimCat detection. We then design an adversarial training procedure with momentum updates and data augmentation to improve robustness of SimCat against adaptive attacks. On ImageNet, the SimCat detector adversarially trained using 25 samples per threat model achieves 71.7\% robustness to a PGD-$\ell_2$ ($\epsilon=2.0$) adaptive attack, a 32\% improvement over vanilla SimCat. Moreover, the adversarially trained SimCat improves the clean accuracy as well, from 73.2\% to 73.6\%. 

In summary, we make the following contributions:
\begin{itemize}
    \item We identify that pre-trained SimCLR embeddings contain valuable information regarding perceptibility of adversarial perturbations. Using this intuition, we develop a sample efficient method for detection and classification of adversarial attacks called SimCat.
    \item We demonstrate the effectiveness of SimCat in detection and classification of various types of adversarial examples in test time (evasion attacks) and training time (poisoning attacks). SimCat leads to impressive empirical results on the ImageNet scale using as few as five training samples per class.
    \item We study adaptive attacks against SimCat and develop an adversarial training procedure that dramatically increases its robustness to adaptive attacks while improving its clean accuracy.
\end{itemize}

\section{Prior Works}

\subsection{Adversarial Attacks and Defenses} 
Given an input $x \in X$ with label $y \in Y$ and a classifier $\mathbf{f}: X \rightarrow Y$, an adversarial attack $\hat{x}$ satisfies $\mathbf{f}(\hat{x}) \not= y, d(\hat{x}-x) \leq \epsilon$, for some generally small bound $\epsilon$. Here, $d(\cdot, \cdot)$ is a distance metric that defines the threat model (i.e. the space of allowable perturbations to the input to craft the adversarial attack). Threat models using $\ell_2$ and $\ell_\infty$ distance are well studied \cite{carlini2017evaluating, madry2018towards}, though attacks that apply spatial transformations, recoloring, frequency domain perturbations \cite{xiao2018spatially,laidlaw2019functional,kang2019robustness} are also effective. Ideally, a defense would ensure that any two images that are imperceptible to a human are classified in the same way. This motivates the neural perceptual threat model of \cite{laidlaw2021perceptual}, which utilizes LPIPS distance as a proxy for the true perceptual distance. 

Defenses against adversarial attacks either focus on robust prediction or detection. Adversarial training \cite{madry2018towards} is the most common method for robust prediction. It operates by crafting and training on adversarial examples, with the ground truth label. While it improves robustness for a specific threat model, the gains do not extend to others. Provable defenses based on smoothing have also been proposed, though they pertain to restricted threat models \cite{DenoisedSmoothing,cohen2019certified}.

Attacks can also be made during training, known as data poisoning \cite{goldblum2020dataset}, where a training set is corrupted so that the trained model misclassifies certain target samples. Clean label poisoning attacks are particularly dangerous and covert, as poisons have the correct label, so the accuracy of the model after training remains high. Two of the strongest clean label poisonings attacks are Convex Polytope (CP) \cite{zhu2019transferable} and Bullseye Polytope (BP) \cite{aghakhani2021bullseye}, which work by making imperceptible changes to a set of baseline images from an intended class so that the features of the baseline images surround a target image, who at test time is then classified to the class of the baseline images. Naturally, robust prediction based defenses can not be applied to data poisoning. 

A number of supervised detection methods have been proposed, based on deep network activations \cite{metzen2017detecting,LSTM}, statistical tests \cite{grosse,LogOdds}, local intrinsic dimensionality \cite{ma2018characterizing}, to name a few. Unsupervised methods based on feature squeezing \cite{FeatureSqueezing}, generative models \cite{PixelDefend}, nearest neighbor search, KL divergence \cite{ADA}, among others, have also been suggested. For a comprehensive review, we refer readers to \cite{Survey}. Generally, unsupervised methods can be costly to configure and sensitive to noise, while supervised methods require lots of data and often fail to generalize to unseen threats. Many detectors have also been shown to be vulnerable \cite{carlini2017evaluating}. To the best of our knowledge, there is no detection system that uses as few samples as SimCat.

Classifying adversarial examples to their respective threat models has been explored in \cite{maini2021perturbation, Liu2020TowardsDM}. This classification can allow for more specific defenses to be applied off the shelf when appropriate, and also give the defender insight about the attacker. 

\begin{figure}
    \centering
    \includegraphics[width=0.75\linewidth]{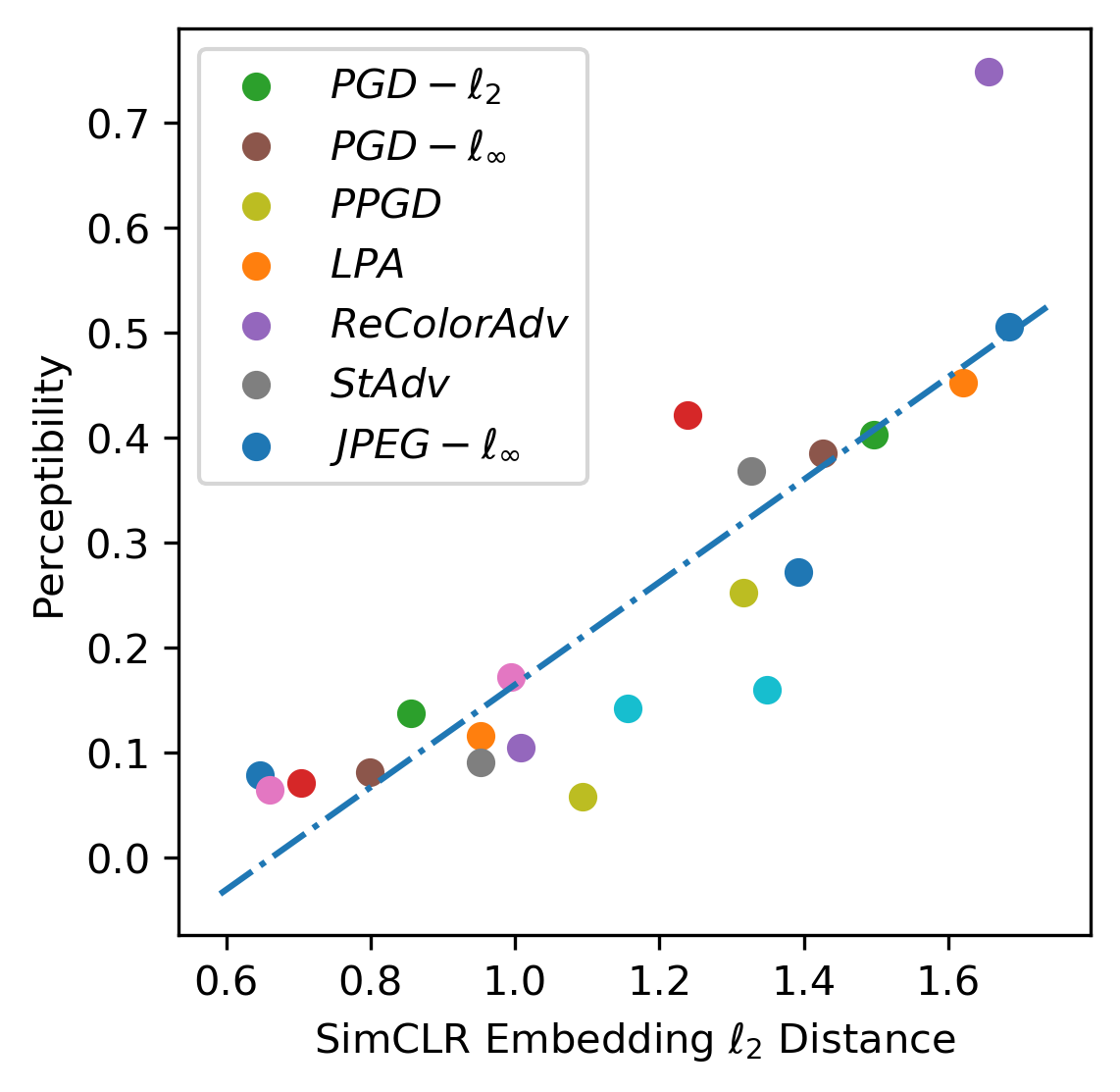}
    \caption{Perceptibility of adversarial attacks relative to the $\ell_2$ distance between the original and perturbed image in the SimCLR embedding space. Each point in the scatter plot refers to the average distance between adversarial examples within a single threat model under one of three bounds. Correlation is $r=0.854$.}
    \label{fig:percep_corr}
\end{figure}

\subsection{Self-Supervised Encoders}

Recent work has seen self-supervised models advance rapidly and in multiple domains \cite{SimCLRv2, CLIP}. We focus on SimCLR \cite{SimCLR}, which is trained using contrastive learning. 

Contrastive Learning is a simple yet powerful self-supervised framework for representation learning that has made large strides in closing the gap with supervised learning. The contrastive loss seeks to maximize similarity between representations of two views of an input, and minimize similarity to views of other samples. SimCLR uses this simple framework, along with a multi-step data augmentation pipeline for generating different views of the same image, to learn very informative visual representations of images. Specifically, SimCLR indirectly applies the contrastive loss on the representations by way of a shallow MLP projection network appended to the encoder during training, and discarded afterwards. While training self-supervised models can be computationally expensive, in our experiments, we use a \textit{fixed} SimCLR encoder \textit{pre-trained} on ImageNet, available openly from \cite{bolts}. 

A few works have looked into adversarially robust contrastive learning \cite{BYORL,CLAE,RoCL,ACL}, though our work differs in that our encoder is fixed and applied to detection and classification of adversarial attacks. Most similar to our work is \cite{sehwag2021ssd}, where SimCLR embeddings are used for anomaly detection. To our knowledge, our work is the first to identify the SimCLR embedding space as one in which adversarial examples and clean images seem to be linearly separable. 
\section{SimCLR distance as proxy for Perceptibility}

\begin{figure*}[h!]
\centering
\begin{minipage}{.5\textwidth}
  \centering
  \includegraphics[width=0.9\linewidth]{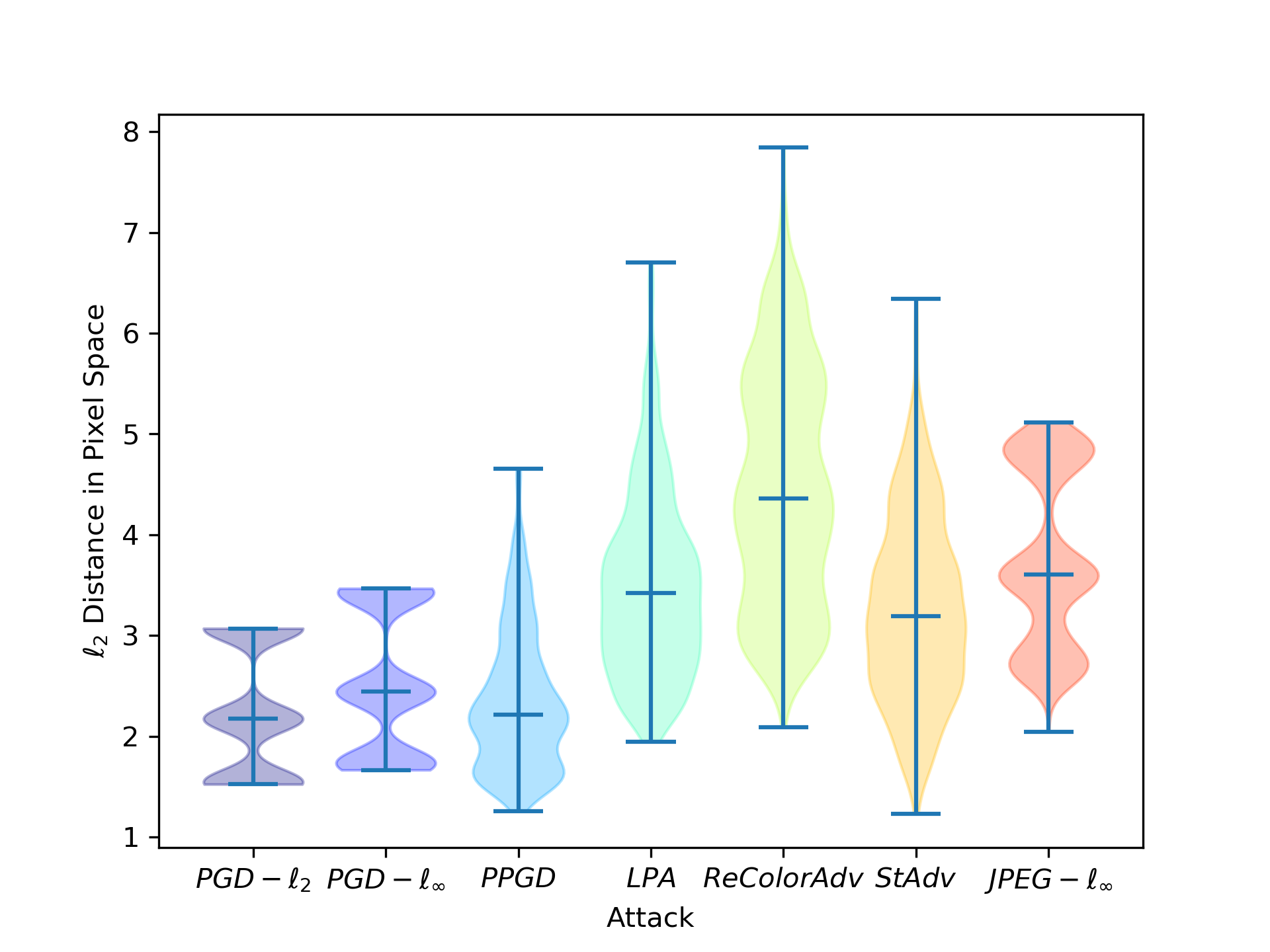}
  \label{fig:test1}
\end{minipage}%
\begin{minipage}{.5\textwidth}
  \centering
  \includegraphics[width=0.9\linewidth]{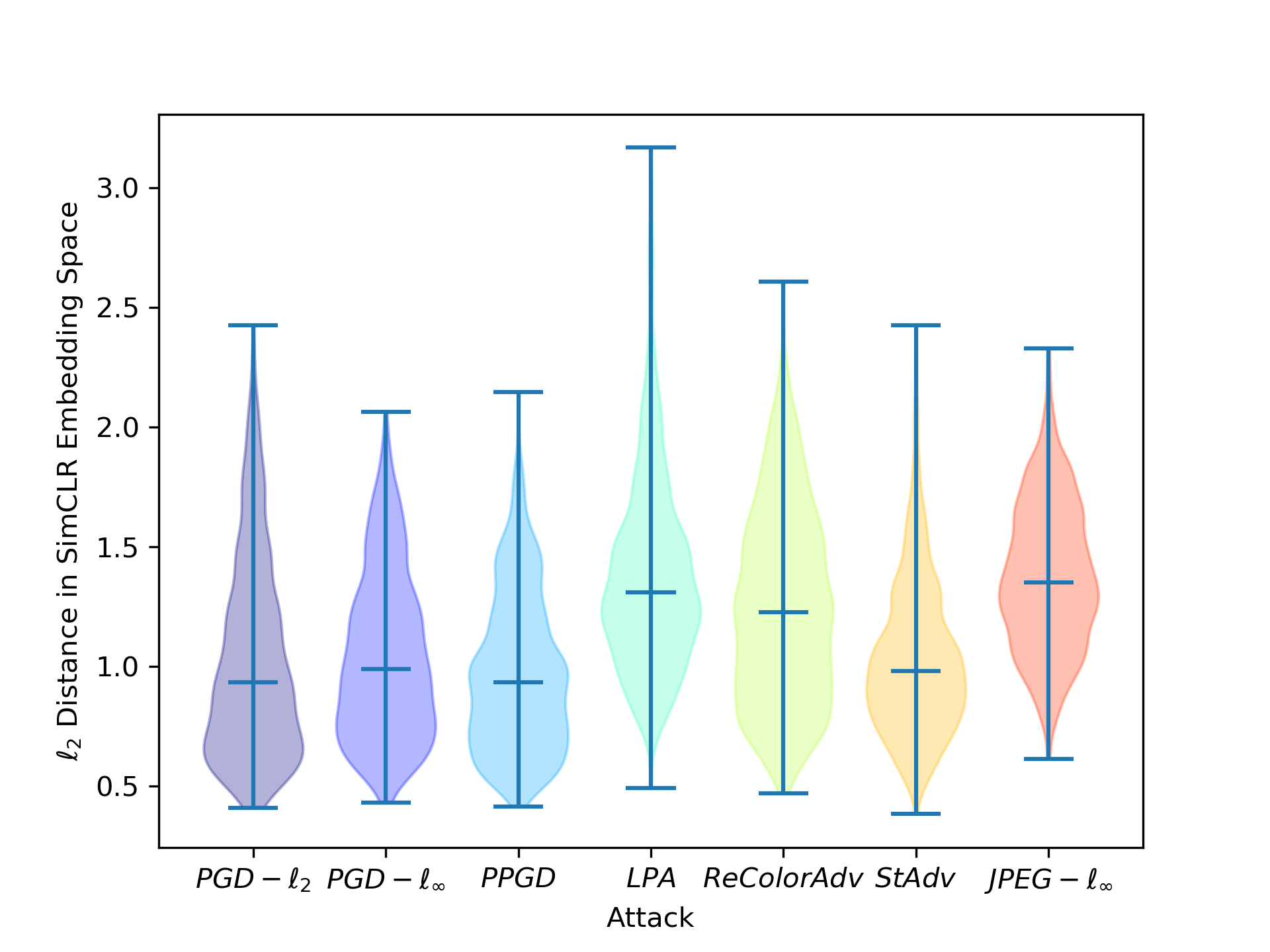}
  \label{fig:test2}
\end{minipage}
\caption{The distributions of $\ell_2$ distances between adversarial examples and clean images in pixel space (left) and SimCLR embedding space (right). Note the different scales; the distributions across threat models are much more uniform when using SimCLR embeddings.}
\label{fig:violins}
\end{figure*}

In this section, we outline the motivation behind using SimCLR as the self-supervised encoder for SimCat. We make use of the data from the human perceptual study in \cite{carlini2017adversarial}. The data consists of seven threat models, spanning perceptual, $\ell_p$, spatial, recoloring, and compression attacks, under three levels of bound on the applied perturbations. Humans were then used to evaluate how perceptible the perturbations were. This was done by presenting a clean and adversarially perturbed sample side by side for two seconds, then having the participant choose whether they thought the images were the same or different. This gives a notion of perceptibly, measured as the ratio of humans who felt the attacked image looked distinct from the original.

In figure $\ref{fig:percep_corr}$, the mean perceptibility over each threat model and attack bound pair is plotted, against the mean $\ell_2$ distance between SimCLR representations of the clean and attacked image. We observe a strong correlation, with Pearson's $r=0.854$. The correlation is reduced by the high perceptibility of the large coloring attacks. This can be explained by the fact that SimCLR is trained to be less sensitive to color shifts, as color jitter is an important augmentation employed in the SimCLR pipeline. Removing the coloring attacks, the correlation improves to $r=0.892$.

Furthermore, SimCLR distance scales similarly for diverse attack types. We observe this in figure $\ref{fig:violins}$, where some non-$\ell_p$ attacks require much higher bounds on $\ell_2$ distance to be encapsulated. On the other hand, SimCLR distance smoothly distributes the attacks of various types and level. This makes the SimCLR distance a strong proxy for the perceptual threat model, suggesting that it could be useful in adversarial training against unseen threat models, though we leave this to a future work. 

We note that the correlation found for $\ell_2$ distance in image space and LPIPS are both comparable to the correlation for the SimCLR distance \cite{laidlaw2021perceptual}. A key advantage of the SimCLR distance over LPIPS is the low dimensionality of its embeddings. While SimCLR only uses a 2048 dimensional representation vector for each input, LPIPS concatenates flattened feature activations from many layers in a deep network to compute distance, which can lead to a blow up in the size of the representation vector, due to the ever increasing depths and widths of modern deep networks (e.g. for LPIPS evaluated on AlexNet, the representation vector has length upwards of 500,000). 

\section{Methods}
\subsection{General SimCat Framework}
In this section, we describe our proposed methods. We use a self-supervised encoder $\phi:\mathbb{R}^D\to\mathbb{R}^d$. Importantly, the self-supervised encoder does not need to be trained or finetuned on the data we wish to apply SimCat to. In our experiments, we use a SimCLR encoder with a ResNet50 backbone pretrained on ImageNet to map inputs into a $d=2048$ dimensional embedding space. Interestingly, the same encoder can be applied effectively to images of varying size (e.g. SVHN (32), STl10 (96), and ImageNet (224)). We apply a linear transformation on the extracted representations to obtain logits. 



For detection, we call our model SimCatch, and denote it by $\mathbf{d_{\boldsymbol{\phi,\omega}}}$, where $\boldsymbol\omega$ contains all the $d+1$ trainable parameters, consisting of vector weights $\mathbf{w} \in \mathbb{R}^d$ and bias $b \in \mathbb{R}$. The SimCatch detector maps $\mathbf{d_{\boldsymbol{\phi,\omega}}}:\mathcal{X} \cup \hat{\mathcal{X}} \rightarrow \mathcal{Y}$, where $\mathcal{X}$ is the space of all natural images, $\hat{\mathcal{X}}$ is the space of all imperceptible adversarial images, and $\mathcal{Y} = \{0,1\}$ is the space of ground truth binary labels, with $1$ denoting an adversarial example. Since it is intractable to capture the entire space of natural and adversarial examples, we estimate $\mathcal{X} \cup \hat{\mathcal{X}}$ with the dataset $D=\bigcup_{i=1}^N \{(\mathbf{x_i}, 0),(\mathbf{\hat{x}_i},1)\}$, where $\hat{\bf x}_i$ is an adversarial example obtained by attacking $\bf x$. An attack specific detector is obtained by restraining the threat model of adversarial examples included in $D$, and an attack agnostic detector seeks to approximate the space of all adversarial examples by sampling from multiple diverse threat models. The output of SimCatch on an input image ${\bf x}$ is

\vspace{-0.1cm}
\begin{equation}
    \text{SimCatch}({\bf x}) := \mathbf{d_{\boldsymbol{\phi, \omega}}}(\mathbf{x}) = sgn(\mathbf{w^T \boldsymbol\phi(x)} + b)
\end{equation}

where \textit{sgn} is the sign function. Note that $D$ does not need to consist of clean and attacked pairs; it can also be two unrelated sets of clean and attacked examples. We hypothesize that training on clean and correspondingly attacked pairs will lead to a more precise decision boundary, but we find in practice that using arbitrary clean samples also suffices.

In classification over $k$ threat models, the vector weights $\mathbf{w}$ are replaced with a matrix $\mathbf{W} \in \mathbb{R}^{k \times d}$.The bias $b$ also now becomes a $d$ dimensional vector $\mathbf{b}$. We refer to this model as SimClass with learnable parameters $\boldsymbol\theta = \{\mathbf{W, b}\}$ and denote it as $\mathbf{g}_{\boldsymbol{\phi, \theta}}: \bigcup_{i=1}^k \mathcal{\hat{X}}_{d_i} \rightarrow [k]$, where $\mathcal{\hat{X}}_{d_i}$ is the space of adversarially perturbed images under a threat model defined by distance metric $d_i$. The training set $D=\bigcup_{i=1}^k\bigcup_{j=1}^N\{(\hat{x}^i_j, i)\}$ consists of $N$ adversarially perturbed examples from each of the $k$ threat models. The output of SimCat used as classifier on input $x$ is then
\begin{equation}
    \text{SimClass}({\bf x}) := \mathbf{g}_{\boldsymbol{\phi, \theta}}({\bf x}) = \argmax_{i\in [k]} (\mathbf{W\boldsymbol\phi(x) + b})_i
\end{equation}

Both SimCat models are trained with a cross entropy loss and $\ell_2$ regularization. Without loss of generality, we present the optimization formulation for SimCatch below. 
\begin{equation}
    \min_{\boldsymbol\omega} \; \sum_{(x,y) \in D}\mathcal{L}_{ce}(\mathbf{d_{\boldsymbol{\phi, \omega}}(x)}, y) + \lambda\|\boldsymbol \omega\|^2
\end{equation}

Importantly, in training, the self-supervised encoder $\boldsymbol \phi$ is \textit{fixed}.
Thus, the number of learnable parameters scales linearly with the number of output classes and dimensionality of the embedding space of $\boldsymbol \phi$. Moreover, the optimization is now \textit{convex}. Due to the low dimensionality of SimCLR's encoder and the convex nature of SimCat's optimization, the global optimum can be found efficiently, in both time and sample complexity. In our experiments, we set the regularization constant $\lambda=1$, and use L-BGFS to obtain optimal parameters for SimCat's regularized logistic regression. 

\subsection{SimCat Variants}

There are many additional modifications that can be made to SimCat to further improve its performance. The majority of the experiments do not use any variants, but in some cases we include the following:
\begin{itemize}
    \item \textbf{Data Augmentation} is a common technique to improve generalizability of deep models. Naturally, data augmentation is very useful when there is limited data available. We utilize data augmentation to balance the dataset during adversarial training (algorithm \ref{alg:at}). Our experiments with data augmentation show improvements in extremely low data settings, though only modest improvement in other cases.
    \item \textbf{Ensembling} can produce improved models by way of combining the outputs of multiple independently trained models. We ensemble detectors for specific poisoning types to improve the performance of our SimCatch based defense (table \ref{tab:poison_defense}). 
\end{itemize}

\section{Detection and Classification}
\subsection{Evasion Attacks}
\subsubsection{Experimental Set up}
\begin{table*}
\begin{center}
\begin{tabular}{|c|c|ccccc|c|}  
\toprule
\multicolumn{2}{|c}{$\textsc{SVHN}$} & \multicolumn{5}{|c|}{Training Samples per Attack} \\ \hline 
Task & Attacks & 2 & 5 & 10 & 25 & 50 \\ \hline \hline
 Detection & PGD-$\ell_2$ & 63.3(+8.8) & 71.9(+11.5) & 75.0(+11.9) &81.7(+13.1)& 85.7(+12.0) \\ \hline
 Detection & PGD-$\ell_\infty$ & 77.3(+15.0) & 82.5(+13.9) & 88.5(+14.0) & 92.4 (+9.9) & 94.2(+8.4) \\ \hline
 Classification & PGD $\ell_2$, PGD $\ell_\infty$ & 60.6(+8.9) & 64.1(+12.8) & 70.9(+16.3) & 77.1(+21.0) & 81.5(+19.6) \\ \toprule 
 \multicolumn{2}{|c}{$\textsc{ImageNet}$} & \multicolumn{5}{|c|}{Training Samples per Attack} \\ \hline 
Task & Attacks & 5 & 10 & 25 & 50 & 100 \\ \hline \hline
Detection &\multirow{4}{7.6em}{PGD-$\ell_2$, PGD-$\ell_\infty$, PPGD, LPA, JPEG-$\ell_\infty$, StAdv, ReColor, CW-$\ell_2$} & 68.5(+6.4)& 71.5(+7.4) & 74.3(+7.4) & 76.5(+5.3) & 79.2(+4.4) \\ 
& & & & & & \\ \cline{1-1}\cline{3-7}
Classification & & 27.1 (+7.7) & 32.8(+10.2) &40.7(+11.8)& 48.9(+12.8)& 58.1(+15.3) \\
& & & & & & \\
\toprule
\end{tabular}
\caption{Performance of SimCat for detection and classification using few training samples on SVHN (top) and ImageNet (bottom). In parenthesis, we denote the improvement gained by using SimCat compared to a baseline using supervised embeddings. For ImageNet, the detector is trained and evaluated over all eight attack types at once, and classification is done over all eight attack types.}
\end{center}
\label{tab:svhn_imagenet}
\end{table*}


We evaluate SimCat's detection and classification capabilities of evasion attacks from two datasets. We compare SimCat's performance to an analagous model that fits a linear layer atop fixed ResNet50 features learned via \textit{supervised} ImageNet pretraining. The baseline differs with SimCat only in that it uses embeddings in a feature space learned with label supervision, highlighting how self-supervised features may better capture distinguishing nuances between the distributions of natural and adversarial images. We present results for a second baseline that additionally finetunes the pretrained ResNet50 in the supplemental material. 

For SVHN \cite{SVHN}, a dataset of street view house numbers with smaller images ($32 \times 32$), we perform PGD $\ell_\infty$ and $\ell_2$ attacks, with budgets of $\epsilon=8/255$ and $\epsilon=1.0$ respectively. We train a detector for each PGD attack, and a classifier to distinguish between the two $\ell_p$ threats. 

For ImageNet, we use the perceptual study data introduced in Section 3. Specifically, we take the attacks of the `large' bound, which have a budget of $\epsilon=8/255$ for the PGD-$\ell_\infty$ attack. The budget for each attack can be found in the original paper. Additionally, we perform Carlini-Wagner-$\ell_2$ attacks on 200 other clean images. We train a \textit{general} detector on all eight attack types, and also a classifier to distinguish between the eight attack types. 

For each adversarial sample, we also have the original clean image. The samples are divided so that the pairs remain in the same set, ensuring that we never have a test image that is either the clean or adversarially perturbed version of a training image. Thus, the total training set size for a detection trial is equal to the number of training samples per attack times 2 times number of attacks, and times $k$ for $k$-way classification. 

In table \ref{tab:svhn_imagenet}, we present results averaged over ten trials, so to account for variability introduced by sampling such a small fraction of the data to train each model.

\subsubsection{Results}
\begin{figure}
    \centering
    \includegraphics[width=0.8\linewidth]{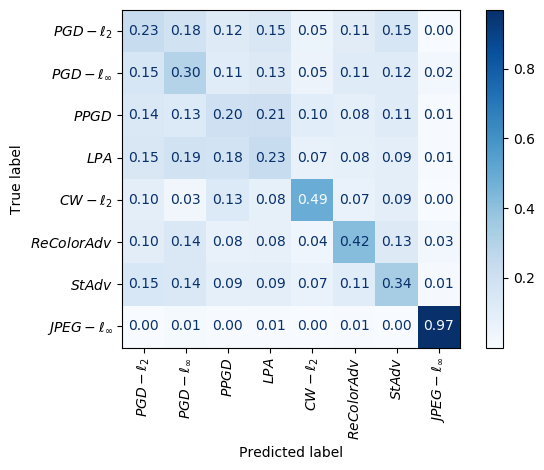}
    \caption{SimCat classification accuracy over eight diverse attack types. The classifier is fit on just 25 samples per class. Overall classification accuracy is 40.7\%, an 11.8\% increase over baseline.}
    \label{fig:classification}
\end{figure}

SimCat outperforms the baseline across the board especially for SVHN, reaching increases in accuracy  of as high as 21.0\%. The efficiency of SimCat is highlighted in SVHN PGD-$\ell_\infty$ detection, where fitting a detector to just two adversarial examples yields 77.3\% accuracy. SimCat's largest gains over the baseline comes in classification tasks (\ref{fig:classification}), indicating that self-supervised features are more sensitive to the distinguishing characteristics of specific attack types. 

\subsection{Poisoning Attacks}
\subsubsection{Experimental Setup}
We test SimCat on five poisoning attacks, including Bullseye Polytope (BP), Convex Polytope (CP), Feature Collision (FC), Clean-label Backdoor (CLBD), and Hidden-trigger Backdoor (HTBD). Poisons are generated using the white-box transfer learning set up as described in \cite{schwarzschild2020just}, where the attacker seeks to poison a fine-tuning set for transfer learning. This setting is generous to the attacker, as they only need to poison a linear layer appended to a fixed feature encoder, that they also have access to. Furthermore, the training set sizes are small. Specifically, the finetuning set only uses 2500 images, and the attacker can insert 1\% (25) additional samples. We use the STL10 dataset, as an intermediary between SVHN and ImageNet. As an additional challenge, we use the SimCLR encoder as the fixed feature encoder -- this means that the poisoning attacks directly cause collisions with clean samples in the same space we wish to use to distinguish them. We test how well clean target samples can be distinguished from poisons that (by design) would be in close proximity to the targets in SimCLR space. We also test SimCat as a poison defense.
\begin{table}[]
\centering
\begin{tabular}{|c|c|c|c|} \bottomrule
\multicolumn{2}{|c|}{Detection} & \multicolumn{2}{|c|}{Classification} \\\hline
Task & Accuracy & Task & Accuracy \\ \hline
BP & 85.3 & Backdoor vs. & 68.9,  78.4$^*$\\ \cline{1-2}
CP  & 84.1&  Triggerless & \\ \hline
General & 64.5, 70.5$^*$& 5-way & 52.4 \\  \toprule
\end{tabular}
\caption{Results for SimCat detection and classification of poisoning attacks on STL10 using ten samples per poisoning. Asterisk indicates removing FC poisons.}
\label{tab:poisons}
\end{table}
\begin{table}[]
\centering
\begin{tabular}{|c|cc|cc|cc|} \bottomrule
\multirow{2}{3em}{Attack Type} & \multicolumn{2}{|c|}{Standard} & \multicolumn{2}{|c|}{SimCat} & \multicolumn{2}{|c|}{Ens. SimCat} \\ \cline{2-7}
& Acc& PSR & Acc & PSR & Acc & PSR \\ \hline
BP & 86 & 82 & 86 & 54 & 85 & 41.3 \\
CP & 86 & 24& 86& 11.3& 85& 4 \\ 
FC & 87 & 0& 86& 1.3& 85&2 \\
CLBD & 87 & 0& 86& 0.7& 85& 0\\
HTBD & 86 & 2& 86& 2& 85& 1.3 \\ \hline
Avg & 86 & 21.8 & 86 & 13.9 & 85 & {\bf 9.7} \\ \toprule
\end{tabular}
\caption{Poison defense via SimCat detection. PSR is poison success rate. SimCat model is trained on CP and BP jointly. Ens. SimCat trains a CP detector and a BP detector separately, then filters any samples that are detected as poison by either detector. Both defenses only use \textit{ten} samples of CP and BP poisons each.}
\label{tab:poison_defense}
\end{table}
\begin{table*}[]
\centering
\begin{tabular}{|c|c|cccccccc|c|} \bottomrule
Trained on: & \# of Samples &PGD-$\ell_2$ &PGD-$\ell_\infty$ & PPGD & LPA &CW-$\ell_2$ &  ReColor & StAdv & JPEG-$\ell_\infty$ & Avg. \\ \hline
Single Attack & 100 & 68.8& 67.9 & 68.5 & 69.2 & 62.2 & 64.9& 65.6 & 50.2 & 64.7\\ \hline
\multirow{2}{3.9em}{Union of Attacks} & 5 & 66.4&  70.7& 65.6&  73.4& 51.5& 71.4 & 63.3 &69.9& 66.5 \\
& 20 &71.1& 76.6& 69.0& 80.1& 51.1& 74.6& 66.1 & 79.4 & 71.1 \\ \toprule
\end{tabular}
\caption{Generalization of SimCat detectors to unseen threat models. First row shows accuracy of detector trained on a single attack evaluated on all other attacks. Other rows contain accuracy of a SimCat detector trained on the union of all other attacks. The second column indicates the number of samples per threat model used in training.}
\label{tab:cv}
\end{table*}
\subsubsection{Results}
SimCat again shows strong detection and classification accuracy with high sample efficiency (\ref{tab:poisons}), particularly for BP and CP poisons, which happen to be the strongest. SimCat struggles with FC poisons, most likely since FC poisons are designed to directly collide with target representations, while BP and CP poisons surround a target instead. When excluding FC poisons, general detection rises to 70.5\%. 

We then apply SimCat detectors as a poison defense. We use BP and CP poisons to train the detectors, since those poisons are most lethal. Using ten samples each, we train an attack agnostic detector, and two separate detectors specific to each threat model, which are used as an ensemble detector, that only admits samples deemed clean by both detectors. Table \ref{tab:poison_defense} shows that both the general detector and the ensemble detector are effective, with the ensemble detector reducing poison success rate from 21.8\% to 9.7\%, while also maintaining high clean accuracy. 

\section{Generalization to Unseen Models}
Generalization of defenses to unseen attacks is of utmost importance because of the constant development of novel threats. In table \ref{tab:cv}, we see that SimCat generalizes surprisingly well given its simplicity. Even when trained only on a single threat model, some of the SimCat detectors achieve close to 70\% detection accuracy on unseen attacks. The generalization of detectors trained on the union of attacks is also impressive, particularly given the sample efficiency. We observe that the detector trained on the union of attacks with just five samples per threat model (35 total) exceeds the average accuracy on unseen attacks achieved by detectors trained on a single threat model with 100 training samples.

\begin{figure}
    \centering
    \includegraphics[width=\linewidth]{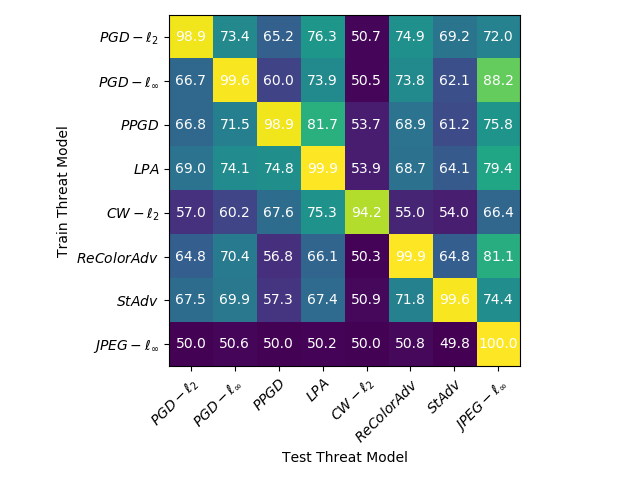}
    \caption{Generalizability of SimCat detectors to unseen threat models. Each detector is trained on 100 samples from a single threat model, and evaluated on all other models.}
    \label{fig:cv_in}
\end{figure}

In figure \ref{fig:cv_in}, we get a closer look at how each threat model generalize to others. The detectors trained on the perceptual attacks (PPGD, LPA) generalize the best to unseen threats. This invokes our motivating observation that the SimCLR embedding space seems to contain information pertinent to perceptibility. Understanding how human and machine perceptions differ is at the heart of many vision tasks, including adversarial robustness, and we encourage future work to further investigate the semantic meaning extracted in SimCLR and other self-supervised models.

\section{Hardening SimCat to an Adaptive Attack}
In this section, we investigate the robustness of SimCat to an adaptive attack. An adaptive attack is an attack that is specifically crafted based on knowledge of a model's defense. By investigating adaptive attacks for SimCat, we can identify limitations of our model, and work towards mitigating them (i.e. via adversarial training) before the vulnerability is exposed and exploited by an adversary. 

\subsection{Attack Formulation}
We consider a white box attack setting, where the attacker has knowledge of the base classifier \textit{and} the SimCat detector. The ultimate goal is to cause a misclassification in the base classifier, but the adversary must first evade the SimCat detector. Denoting the base classifier as $\mathbf{f}$, the detector as $\mathbf{d}$, and an input-label pair as $(x, y)$, we formulate the adaptive adversarial attack problem as the following.
\begin{equation}
    \boldsymbol\delta = \argmax_{\boldsymbol\delta, \|\boldsymbol\delta\|\leq \epsilon} \mathcal{L}(\mathbf{f}({\bf \hat{x}}+\boldsymbol\delta), y) + \mathcal{L}(\mathbf{d}_{\boldsymbol{\phi,\omega}}({\bf \hat{x}}+\boldsymbol\delta), 1)
\end{equation}

For both terms, $\mathcal{L}$ is the cross entropy loss. The detector outputs $1$ for adversarial examples, so the adaptive attack seeks to flip this label by maximizing the loss incurred by it. We solve the above optimization with projected gradient descent, and find that the adaptive attack is somewhat effective against an undefended SimCat ImageNet detector, reducing accuracy by 30\% (table \ref{tab:adaptive}).

\subsection{Adversarial Training}
We employ adversarial training (AT) to improve the robustness of SimCat to the adaptive attacks described in the previous section. Standard AT seeks to harden a network by crafting adversarial examples throughout training, and additionally training the model on the crafted examples with the original label. For SimCat, this amounts to the following min-max optimization, where $\mathbf{d}$ is the SimCat detector with parameters $\boldsymbol{\phi, \omega}$, and $\mathbf{f}$ is the base classifier.
\begin{equation}
    \min_{\boldsymbol \omega} \max_{\boldsymbol\delta; \|\boldsymbol\delta\| \leq \epsilon} \mathcal{L}_{ce}({\bf f}( {\bf \hat{x}}+\boldsymbol\delta), y) + \lambda \mathcal{L}_{ce}({\bf d}_{\boldsymbol{\phi, \omega}}({\bf \hat{x}}+\boldsymbol\delta), 1)
\end{equation}

\begin{figure}
    \centering
    \includegraphics[width=0.85\linewidth]{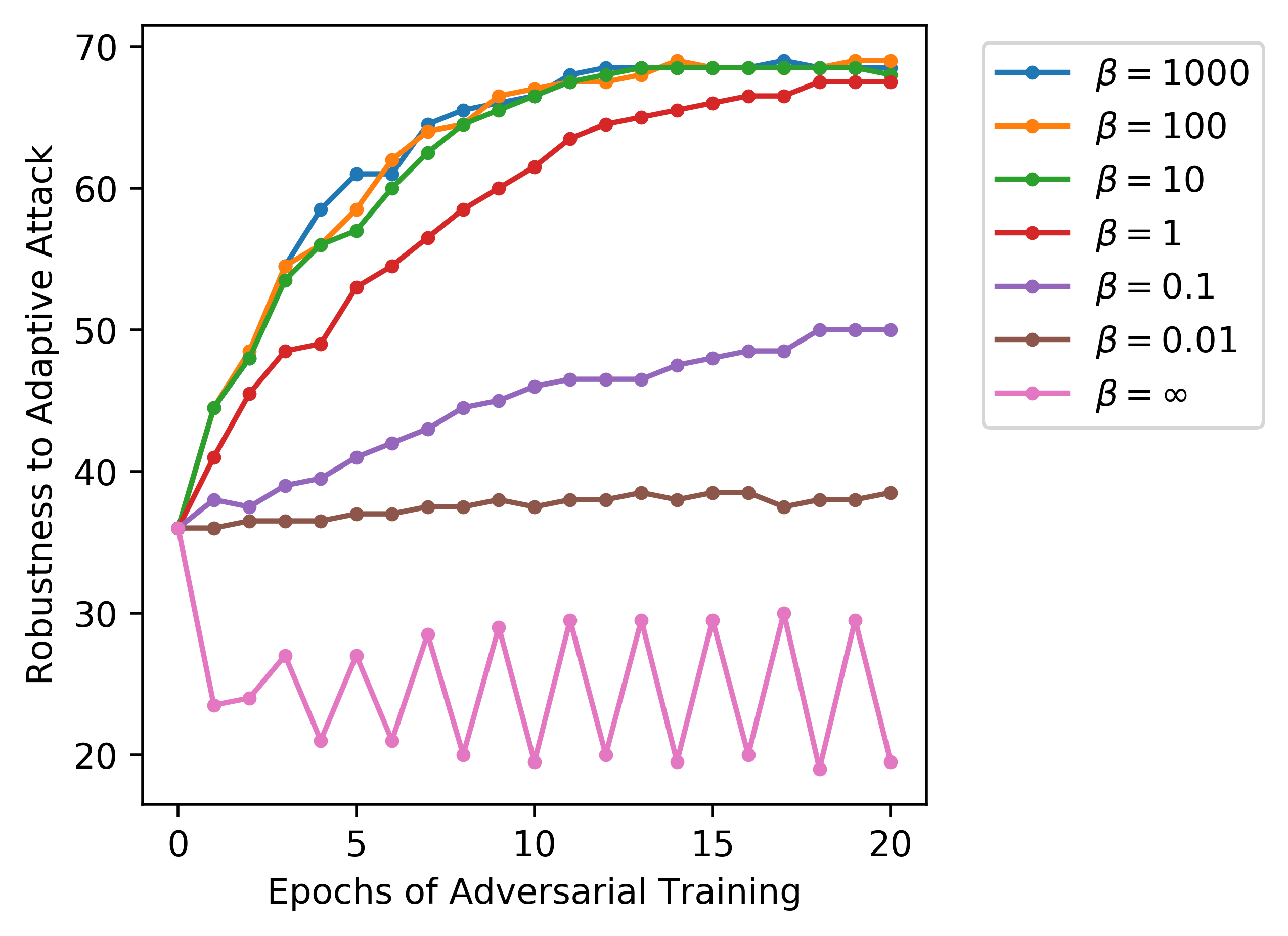}
    \caption{SimCat robustness to adaptive PGD-$\ell_2$ over epochs of adversarial training with varied values of the hyperparameter $\beta$, which controls the momentum updates. Higher values of $\beta$ lead to more emphasis on the linear classifiers solved in later epochs. Adversarial training is unstable without momentum ($\beta=\infty$).}
    \label{fig:varying_beta}
    \vspace{-0.45cm}
\end{figure}

SimCat AT is different from standard AT in a few ways. Standard AT usually takes a few steps of finding perturbations to increase the objective, followed by a few steps of updating model parameters to reduce the objective. In SimCat AT, the minimization step is solved to completion after having crafted adaptive adversarial examples, as opposed to only taking a few steps. This can be done efficiently due to the low number of parameters to solve for and the convexity of the minimization problem. A couple other steps are needed for SimCat AT to be effective.
\begin{itemize}
    \item Momentum updates are used to stabilize training. The importance of momentum updates is clear in figure $\ref{fig:varying_beta}$, as the $\beta=\infty$ case (where the SimCat detector is replaced in each epoch with the optimal solution for detecting the current batch of adaptive adversarial examples) yields worse robustness than a standard SimCat. 
    \item Along with the additional adaptive adversarial attacks, the original data is retained in the training set for each iteration, so to mitigate a robustness-accuracy tradeoff. To balance the dataset, an augmented copy of the clean samples is also added to the training set in each epoch.  
\end{itemize}
\vspace{-0.1cm}
\begin{algorithm}
\caption{Adversarial Training of SimCat: inputs are the base classifier $\mathbf{f}$ and data $\left(\{(x_i, \hat{x}_i)\}_{i=1}^N\right)$, where $x$ is a clean sample and $\hat{x}$ is $x$ after being adversarially perturbed.}
\begin{algorithmic}
\STATE Obtain initial parameters via standard SimCat:
\STATE $\boldsymbol\omega \leftarrow \textsc{FitSimCat}\left(\{(x_i, \hat{x}_i)\}_{i=1}^N\right)$
\STATE Augment clean data to obtain second copy:
\STATE $\{\tilde{x}_i\}_{i=1}^N \leftarrow \textsc{Augment}\left(\{x_i\}_{i=1}^N\right)$
\FOR{$t =1, \dots, \text{\# of epochs}$}
\STATE Craft perturbations for adversarial examples to evade both detector and base classifier: 
\STATE $\{\delta_{i,t}\}_{i=1}^N = \textsc{AdaptivePGD}\left(\{\hat{x}_i\}_{i=1}^N, \mathbf{f}, \mathbf{d_{\boldsymbol{\phi, \omega}}}\right)$
\STATE Solve SimCat with expanded dataset:
\STATE $\boldsymbol\omega_t \leftarrow \textsc{FitSimCat}\left(\{(x_i, \hat{x}_i), (\tilde{x}_i, \hat{x}_i+\delta_{i,t})\}_{i=1}^N \right)$
\STATE Apply momentum update to SimCat parameters:
\STATE $\boldsymbol\omega \leftarrow (\boldsymbol\omega+\beta\boldsymbol\omega_t) / (1+\beta)$
\ENDFOR
\end{algorithmic}
\label{alg:at}
\end{algorithm}
\vspace{-0.1cm}
\subsection{Results}
\begin{table}[]
\centering
\begin{tabular}{|c|c|c|}\bottomrule
 Model & Accuracy & Robustness \\ \hline
 SimCat & 73.21 &  39.25 \\
 SimCat+Aug & 74.87 & 37.40 \\
 SimCat+AT & 74.23 & 67.95\\
 SimCat+AT+Aug & {\bf 73.55} & {\bf 71.70}\\ \toprule
\end{tabular}
\caption{Ablation study on SimCat AT. Accuracy refers to attack agnostic ImageNet detection from Section 5.1. Robustness is measured as the percent of test adversarial samples that can be adaptively attacked with PGD-$\ell_2$, $\epsilon=2.0$ to be misdetected as clean. AT is done for 20 epochs. Aug refers to augmenting both the original clean and adversarial samples - this is distinct from AT+Aug, where only the clean samples are augmented to balance out the addition of adaptive adversarial attacks to the SimCat training set.}
\label{tab:adaptive}
\vspace{-0.15cm}
\end{table}

While the adaptive adversarial attack is somewhat effective (reducing attack detection accuracy by 34\%), the SimCat AT algorithm completely recovers robust accuracy, while also improving overall accuracy. Table $\ref{tab:adaptive}$ shows the effect of AT and augmentation, which in tandem become a very strong defense. Furthermore, the entire adversarial training procedure takes only about fifteen minutes, and all SimCat AT model results presented used only 25 training samples per attack. Thus, the SimCat framework lends itself to increased robustness via algorithm \ref{alg:at}, without compromising training and data efficiency.

\section{Conclusion}
In this paper, we introduced SimCat, a sample efficient method for detection and classification of adversarial attacks. SimCat uses a linear model over embeddings of a pre-trained self-supervised model, SimCLR. SimCat is successful in detecting and classifying various types of adversarial attacks ranging from $\ell_p$ and non-$\ell_p$ evasion attacks to poisining attacks, likely because pre-trained SimCLR embeddings can be used to uniformly quantify perceptibility of various types of adversarial perturbations. Over various experiments on SVHN, ImageNet, and STL10 datasets, we demonstrate the effectiveness of SimCat using as few as two training samples per class. We have also studied adaptive attacks against SimCat and developed an adversarial training procedure that dramatically increases its robustness against such attacks while improving its clean accuracy.

\section{Acknowledgements}
This project was supported in part by NSF CAREER AWARD 1942230, HR00112090132, HR001119S0026 and ONR GRANT13370299.

\newpage

{\small
\bibliographystyle{ieee_fullname}
\bibliography{egbib}
}

\end{document}


\title{Supplementary Materials}

\author{Mazda Moayeri\\
Department of Computer Science \\
University of Maryland\\
{\tt\small mmoayeri@umd.edu}
\and
Soheil Feizi\\
Department of Computer Science \\
University of Maryland\\
{\tt\small sfeizi@cs.umd.edu}
}

\maketitle


\section{Training Details}

\begin{table}[]
    \centering
    \begin{tabular}{c|c} \toprule
        \multicolumn{2}{c}{Hyperparameter Details} \\ \midrule \midrule
        \multicolumn{2}{c}{Baseline Finetuning} \\ \midrule
        Architecture & ResNet50 \\
        Optimizer & SGD \\
        Learning Rate & 0.001 \\
        Momentum & 0.9 \\
        Epochs & 50 \\\midrule
        \multicolumn{2}{c}{SimCLR Encoder Pretraining$^*$} \\ \midrule
        Architecture & ResNet50 \\ 
        Dataset & ImageNet \\
        Optimizer & LARS-SGD \\
        Learning Rate & 4.8 \\
        Batch size & 4096 \\
        Epochs & 800 \\ \midrule
        \multicolumn{2}{c}{Logistic Regression} \\ \midrule
        Optimizer & LGBFS \\
        Max Iterations & 1500 \\
        Tolerance & 0.001 \\
        Regularization Penalty & $\ell_2$ \\
        Regularization Penalty Weight & 1 \\ \bottomrule
    \end{tabular}
    \caption{Various hyperparameter choices for experiments presented in main text. *Note that SimCLR encoder was downloaded from \cite{bolts} {\it already trained}. We report some details of their training procedure, as presented on their API.}
    \label{tab:hyperparams}
\end{table}

In this section, we provide additional details to the training of the models presented in the main text. The information is also presented in Table \ref{tab:hyperparams}. 

\subsection{Baseline Training}

\begin{table*}[h]
\centering
\begin{tabular}{|c|c|ccccc|c|}  
\toprule
\multicolumn{2}{|c}{$\textsc{SVHN}$} & \multicolumn{5}{|c|}{Training Samples per Attack} \\ \hline 
Task & Attacks & 2 & 5 & 10 & 25 & 50 \\ \hline \hline
 Detection & PGD-$\ell_2$ & 63.3 (+13.0) & 71.9 (+19.3) & 75.0 (+17.0) &81.7 (+7.6)& 85.7 (+19.2) \\ \hline
 Detection & PGD-$\ell_\infty$ & 77.3 (+25.1) & 82.5 (+25.5) & 88.5 (+21.2) & 92.4 (+6.9) & 94.2 (+2) \\ \hline
 Classification & PGD $\ell_2$, PGD $\ell_\infty$ & 60.6 (+9.9) & 64.1 (+9.4) & 70.9 (+11.7) & 77.1 (+5.5) & 81.5 (+19.6) \\ \toprule 
 \multicolumn{2}{|c}{$\textsc{ImageNet}$} & \multicolumn{5}{|c|}{Training Samples per Attack} \\ \hline 
Task & Attacks & 5 & 10 & 25 & 50 & 100 \\ \hline \hline
Detection &\multirow{4}{7.6em}{PGD-$\ell_2$, PGD-$\ell_\infty$, PPGD, LPA, JPEG-$\ell_\infty$, StAdv, ReColor, CW-$\ell_2$} & 68.5 (+0.7)& 71.5 (+3.4) & 74.3 (+4.3) & 76.5 (+3.4) & 79.2 (+2.8) \\ 
& & & & & & \\ \cline{1-1}\cline{3-7}
Classification & & 27.1 (+3.0) & 32.8 (+1.5) &40.7 (+0.9) & 48.9 (-1.9) & 58.1 (-1.9) \\
& & & & & & \\
\toprule
\end{tabular}
\label{tab:baseline}
\caption{Performance of SimCat for detection and classification using few training samples on SVHN (top) and ImageNet (bottom). In parenthesis, we denote the improvement gained by using SimCat compared to a baseline finetuned pretrained Resnet50 classifier. For ImageNet, the detector is trained and evaluated over all eight attack types at once, and classification is done over all eight attack types.}
\end{table*}

We train two baseline models. The primary baseline model (with results presented in the main text) is designed to be as similar to SimCat as possible. The sole difference is that the baseline model obtains embeddings by taking activations from the penultimate layer of a ResNet50 network pretrained on ImageNet in standard supervised fashion. Thus, the convex optimization, backbone, and pretrained data are exactly the same between the baseline and SimCat. The intention with this baseline is to highlight the way in which the distributions of clean and adversarially attacked images organize differently in supervised and self supervised embedding spaces.

The second baseline allows for further fine-tuning of the pretrained ResNet50. Fine tuning is common practice when data is limited, and we believe using a pretrained ResNet highlights how features learned in a self-supervised fashion are distinct from those learned with label supervision, such that they better separate attacked and clean images. 

Note that choosing the best state for the fine tuned baseline is challenging in the low data setting, as there is no hold out set to evaluate generalizability. The reported results for the fine tuned baseline average the test accuracy for the final ten epochs of baseline training.

\subsection{SimCat Training}

The SimCat models consist of a fixed SimCLR encoder, with a linear layer appended on top for either adversarial attack classification or detection. The training of SimCat has two steps: the self-supervised training of the encoder using the contrastive loss, and the convex optimization of the linear layer. 

Importantly, we stress that the we do not train the SimCLR model, and instead we obtain a pretrained model from \cite{bolts}. The pretrained SimCLR encoder we use in our experiments has a ResNet50 backbone. It was trained for 800 epochs on ImageNet data, with a batch size of 4096 and a LARS-SGD optimizer with a learning rate of 4.8. 

We solve the logistic regression for fitting the linear component of SimCat with an LGBFS optimizer for 1500 iterations, or to completion, with a tolerance of 0.001 (all SimCat models converged to an error under the tolerance within the 1500 iterations). An $\ell_2$ regularization with penalty weight $1$ is applied. 

\section{Fine Tuned Baseline Results}

\begin{table}[h]
    \centering
    \begin{tabular}{c|cc}
        \toprule
         \textsc{SVHN} & Fixed+Linear & Fine Tuned \\ \midrule
         Detection PGD-$\ell_2$ & +11.5 & +15.2 \\
         Detection PGD-$\ell_\infty$ & +12.2 & +16.2 \\
         Classification (2-way) & +15.7 & +11.2 \\ \toprule
         \textsc{ImageNet} & Fixed+Linear & Fine Tuned \\ \midrule
         Multi-attack Detection & +6.2 & +2.9 \\
         Classification (8-way) & +11.6 & +0.3 \\ \toprule
    \end{tabular}
    \caption{SimCat accuracy gain over two new baselines: i. fitting a linear layer atop fixed ResNet50 features (Fixed + Linear); ii. fine tuning a ResNet50 (Fine Tuned). Both use a ResNet50 {\it pretrained on ImageNet}, making them stronger than our original baseline (trained from scratch). Gains are averaged over all training set sizes originally studied.}
    \label{tab:baseline_avg}
\end{table}

In this section, we present SimCat's accuracies for evasion attack detection and classification relative to the fine-tuned baseline in Table \ref{tab:baseline}, and summarize the gain over each baseline in Table \ref{tab:baseline_avg}. 

Both baselines perform much more competitively to SimCat for the tasks on ImageNet data, while the gap is significantly larger for tasks on SVHN data. This could potentially be caused by the fact that the data seen during pretraining was from ImageNet, suggesting that when it comes to capturing the nuances in the distribution of adversarial examples, feature spaces learned in a supervised fashion may be more sensitive to changes in data sets than self-supervised embeddings. This prompts future work investigating the out-of-distribution robustness of self-supervised features relative to supervised features. 

The fine tuned baseline only seems to consistently outperform the other baseline in the Imagenet tasks. In general, the ImageNet tasks are more challenging, because of the eight attack types considered. This could explain the improvement of the fine tuned baseline, as the network then has much more expressive power to approximate a potentially more complicated decision boundary. However, there are some instances where increasing the number of training samples per attack from 25 to 50 cause a significant drop off in performance of the fine tuned baseline (SVHN PGD-$\ell_2$ detection, SVHN PGD-$\ell_2$ vs PGD-$\ell_\infty$ classification), highlighting the vulnerability of overfitting in supervised models trained on small amounts of data. Seeing as both SimCat and the fixed+linear baseline do not experience similar drop offs, it is possible that the simple model complexity of a single learnable linear layer and the convex optimization of regularized logistic regression creates increased generalizability. 

\section{Generating Adversarial Examples}

\subsection{Evasion Attacks}

\begin{table}[]
    \centering
    \begin{tabular}{|c|c|c|c|} \hline
         Attack & Bound & Step size & Iterations  \\ \hline
         PGD-$\ell_2$ & 1.0 & 0.2 & 40 \\
         PGD-$\ell_\infty$ & 8 / 255 & 2 / 255 & 40 \\ \hline
    \end{tabular}
    \caption{SVHN PGD attack hyperparameters.}
    \label{tab:svhn_pgd}
\end{table}

We experiment on two sets of evasion attacks. For each threat model, we obtain 200 poisons. The first set comes from performing $\ell_2$ and $\ell_\infty$ PGD attacks on SVHN images, with budgets of $\epsilon=1.0$ and $\epsilon=8/255$ respectively. Full details on the attacks are presented in table \ref{tab:svhn_pgd}. 

\begin{table}[]
    \centering
    \begin{tabular}{c|c} \bottomrule
         Attack & Bound   \\ \hline
         PGD-$\ell_2$ & 2400 \\
         PGD-$\ell_\infty$ & 8 \\
         PPGD & 1 \\
         LPA & 1 \\
         JPEG-$\ell_\infty$& 0.25 \\
         ReColorAdv & 0.12 \\
         StAdv & 0.1 \\
         CW $\ell_2$ & 0.25 \\ \toprule
    \end{tabular}
    \caption{Bounds on each attack for the ImageNet dataset, as obtained from the user study data from \cite{laidlaw2021perceptual}. The entry for Carlini-Wagner attacks refers to the $c$ value in the attack formulation, which controls strength of attack relative to amount of distortion. The bounds assume images have pixels in the range [0,255], though this only effects the entries for PGD-$\ell_2$, PGD-$\ell_\infty$, and JPEG-$\ell_\infty$ attacks.}
    \label{tab:imagenet_bounds}
\end{table}

The ImageNet adversarial attacks were obtained from the human study conducted in \cite{laidlaw2021perceptual}. We use the data from the `large' level bound for each of the attacks. The specific bounds for each threat model is described in table \ref{tab:imagenet_bounds}, as informed by the values in appendix D in \cite{laidlaw2021perceptual}, which also contains the attack bounds for the `small' and `medium' level attacks, whose perceptibilities are presented in Section 3 of the main text. Additionally, we perform Carlini-Wagner $\ell_2$ attacks on 200 images. One important distinction is that we set $c=0.25$, while the original Carlini-Wagner formulation finds (via binary search) and uses the smallest $c$ value that yields a successful attack. We choose the larger $c$ value to match the choice of studying attacks from the `large' bound. 

\subsection{Poisoning Attacks}

Poisons of 5 different types were constructed, including Bullseye Polytope, Convex Polytope, Feature Collision, Clean Label Backdoor, and Hidden Trigger Backdoor, all following the benchmark protocol outlined in \cite{schwarzschild2020just} and using their publicly available implementation of the benchmark, specifically for the white box transfer learning setting. The only modifications we make to the benchmark is to use STL10 data and the pretrained SimCLR encoder as the fixed feature extractor, on which poisons are developed. The choice of using the pretrained SimCLR encoder to generate poisons was to see if SimCat would be vulnerable to poisons who are generated specifically against it. In table \ref{tab:poison_gen}, we list relevant hyperparameters for the poison generation. We refer the readers to the benchmark for more detail on the procedure for generating each poison. 

\begin{table}[]
    \centering
    \begin{tabular}{c|c} \toprule
    \multicolumn{2}{c}{All Poisons} \\ \midrule
    Norm Constraint & $\ell_\infty$ \\
    Perturbation Bound & 8 / 255 \\ 
    Dataset & STL10 \\
    Image Size & $96 \times 96$ \\
    Number of Poisons per target & 25 \\
    Number of Targets & 100 \\ \midrule
    \multicolumn{2}{c}{Bullseye Polytope} \\ \midrule
    Crafting Iterations &  1200 \\
    Learning Rate & 0.04 \\
    Optimizer & Adam, Betas (0.9, 0.999) \\ \midrule
    \multicolumn{2}{c}{Convex Polytope} \\ \midrule
    Crafting Iterations &  1200 \\
    Learning Rate & 0.04 \\
    Optimizer & Adam, Betas (0.9, 0.999) \\ \midrule
    \multicolumn{2}{c}{Feature Collision} \\ \midrule
    Crafting Iterations &  120 \\
    Step Size & 0.001 \\
    Watermark Coefficient & 0.3 \\ \midrule
    \multicolumn{2}{c}{Clean Label Backdoor} \\ \midrule
    Number of Steps & 20 \\
    Step size & 2 / 255 \\
    Patch size & 15 \\ \midrule
    \multicolumn{2}{c}{Hidden Trigger Backdoor} \\ \midrule
    Crafting Iterations & 5000 \\
    Learning Rate & 0.001 \\
    Patch size & 15 \\ \bottomrule
    \end{tabular}
    \caption{Details on poison generation. All hyperparameters follow the choices set in \cite{schwarzschild2020just}.} 
    \label{tab:poison_gen}
\end{table}

For each poison type, 100 target-poisons sets are crafted, consisting of a single target and 25 corresponding poisons. During poison defense, we reserve the first 50 sets for training of detectors, and evaluate poisonings on the latter 50 sets. Note that we only use a small subset of the reserved poisons in order to train each detector.

\section{Evaluating Poisonings and Poison Defense}

\subsection{Transfer Learning Benchmark Overview}
Following the white box transfer learning setting described in the benchmark, in each poisoning attempt, a linear classifier is trained on the representations of 2500 clean samples and 25 poison samples, as obtained by the fixed feature encoder that was to used generate poisons (hence white box). As a reminder, in our experiments, the fixed feature encoder is the pretrained SimCLR encoder, and the 2500 clean samples are obtained by taking the first 250 samples for each class in the STL10 training set. A poisoning attempt is successful if the target is misclassified to the desired class (i.e. the class of the poisons). Poison success rate is the ratio of successful poisoning attempts out of the 50 total possible successes. We also report the clean accuracy over all 8000 test samples achieved by the finetuned model. 

\subsection{Applying SimCat Detector}

In a single trial for poison defense, we train three SimCat detectors: one on only Bullseye Polytope poisons, one on only Convex Polytope poisons, and a general detector on both. The two poison specific detectors are combined to form an ensemble SimCat detector (denoted Simcat+Ens in Table 3 in the main text). The training set for each poison-specific detector is constructed by first selecting ten targets randomly from the 50 targets reserved for training per poison type. Then, for each target, one out of the 25 corresponding poisons is randomly selected and added to the training set, yielding a total of twenty samples to fit the SimCat detector over. An ensemble detector is constructed by only marking an image as clean if both poison-specific detectors mark the image as clean. 

For the general detector, we accumulate the same randomly selected poisons, but use other clean samples, so to avoid double counting the same target images in the training set. Thus, the training set for the ensemble detector and the general detector in a given trial uses the same poisons, but different clean samples. 

Then, to evaluate a detector (general or ensemble), we apply it over the entirety of the finetuning set and all poisons to be considered, removing any samples marked as poison. Then, we evaluate each poisoning, inserting only poisons that bypass the detector, and also only using the clean samples that were not falsely marked as poison by the detector. In total, there are 50 poisonings tested for each poison type. 

We repeat this process over five trials, reporting the average results in Table 3 in the main text. 

\section{ImageNet Evasion Attack Classification}

In this section, we offer additional results for ImageNet adversarial attack classification. In table \ref{tab:classification_splits}, we show the corresponding heat maps for SimCat classification of eight evasion attack types at varying levels of sample efficiency, corresponding to the results in Table 1 in the main text. All results presented are averaged over ten trials. While SimCat achieves reasonable accuracies with as little as 10 training samples per attack type, the performance increases dramatically when allowing for as many as 100 training samples per attack type. We note that this is still too small a training set to successfully train a baseline classifier. However, unlike with detection, the baseline classifier largely exceeds random guessing, outperforming random classification by about a factor of 2.5. 

\begin{table}[]
    \centering
    \begin{tabular}{c|c} \toprule 
    \multicolumn{2}{c}{Adaptive Attack} \\ \midrule
        Norm Constraint & $\ell_2$ \\
        Perturbation Bound & 2.0 \\
        Attack method & PGD \\
        Number of Steps & 20 \\
        Step Size & 0.05 \\ \midrule
    \multicolumn{2}{c}{Adversarial Training} \\ \midrule
        Epochs & 20 \\
        Training Samples per Attack & 25 \\
        $\beta$ & 100 \\
         & Random Resized\\
        Augmentation & Crop, then Random \\
        & Horizontal Flip \\ \bottomrule
    \end{tabular}
    \caption{Details on hyperparameters for adaptive attack and adversarial training.}
    \label{tab:adaptive_dets}
\end{table}
\begin{table*}[]
    \centering
    \begin{tabular}{|c|c|} \hline
    10 samples per threat model & 25 samples per threat model \\
    Accuracy: 32.8 & Accuracy: 40.7 \\
    \includegraphics[width=0.48\linewidth]{LaTeX/cm_10_samples_each_class.png} & \includegraphics[width=0.48\linewidth]{LaTeX/cm_25_samples_each_class.png}  \\ \hline
    50 samples per threat model & 100 samples per threat model \\ 
    Accuracy: 48.9 & Accuracy: 58.1\\
    \includegraphics[width=0.48\linewidth]{LaTeX/cm_50_samples_each_class.png} & \includegraphics[width=0.48\linewidth]{LaTeX/cm_100_samples_each_class.png} \\ \hline
    \end{tabular}
    \caption{SimCat classification accuracy of eight attack types over different sizes of training sets. Each figure is labeled with the number of samples per threat model included in the training set, as well as the classification accuracy. All results are averaged over ten trials.}
    \label{tab:classification_splits}
\end{table*}

\section{Adaptive Adversarial Training}

In this section, we elaborate on the experimental procedure for the adaptive attack and adversarial training mentioned in Section 7 of the main text. In table \ref{tab:adaptive_dets}, we provide the details for the attack used during adversarial training, as well as details on adversarial training results found in Table 5 of the main text.

A clarifying note is that in adversarial training, we craft adaptive attacks by designing perturbations to the \textit{adversarial examples} in our training set. This way, we do not simply perform adversarial training on PGD-$\ell_2$ attacks, but instead train over the diverse set of adversarial attacks in our training set, as well as slightly perturbed versions of those attacks, such that they also evade detection. This is made clear in the algorithm pseudocode in the main text, but may otherwise by ambiguous. To further attempt to retain the qualities of the different attack types present in our training set, we choose a small step size and low number of PGD steps in our attack.

{\small
\bibliographystyle{ieee_fullname}
\bibliography{egbib}
}